\documentclass[aps,prl,twocolumn,showpacs]{revtex4}
\usepackage{amssymb,amsmath}
\usepackage[dvips]{graphicx}

\newcommand{\bd}[1]{\mbox{\boldmath$#1$}}

\begin{document}

\title{From the Kuramoto-Sakaguchi model to the Kuramoto-Sivashinsky equation}

\author{Yoji Kawamura}
\email{ykawamura@jamstec.go.jp}
\affiliation{Institute for Research on Earth Evolution,
Japan Agency for Marine-Earth Science and Technology, Yokohama 236-0001, Japan}


\date{January 14, 2014}    


\begin{abstract}
  We derive the Kuramoto-Sivashinsky-type phase equation
  from the Kuramoto-Sakaguchi-type phase model
  via the Ott-Antonsen-type complex amplitude equation
  and demonstrate heterogeneity-induced collective-phase turbulence
  in nonlocally coupled individual-phase oscillators.
\end{abstract}

\maketitle


\paragraph{Introduction.}

Large populations of coupled limit-cycle oscillators exhibit various types of collective behavior
\cite{ref:winfree80,ref:kuramoto84,ref:pikovsky01}.
Among them,
the following two types of collective dynamics emerging from a system of coupled oscillators
have received considerable attention:
collective synchronization in globally coupled systems
\cite{ref:kuramoto75,ref:sakaguchi86,ref:kuramoto02,ref:strogatz00,ref:acebron05,ref:arenas08,
  ref:kiss02,ref:kiss07,ref:taylor09,ref:tinsley12,ref:nkomo13}
and
pattern formation including spatiotemporal chaos in locally coupled systems
\cite{ref:kuramoto74,ref:kuramoto76,ref:cross93,ref:aranson02,
  ref:kim01,ref:mikhailov06,ref:mikhailov13}.
The phase description method~\cite{ref:winfree80,ref:kuramoto84,ref:pikovsky01},
which enables us to describe the dynamics of an oscillator by a single variable called the phase,
is commonly used to analyze the system of coupled oscillators.
On one hand,
a phase description approach to collective synchronization
resulted in the Kuramoto model~\cite{ref:kuramoto84,ref:kuramoto75},
which was generalized
by including a phase shift
to the Kuramoto-Sakaguchi model~\cite{ref:sakaguchi86}:
\begin{equation}
  \dot{\phi}_j(t)
  = \omega_j - \frac{K}{N} \sum_{k=1}^N \sin(\phi_j - \phi_k + \alpha).
  \label{eq:Kuramoto-Sakaguchi_model}
\end{equation}
This type of phase model has been experimentally realized
using electrochemical oscillators~\cite{ref:kiss02,ref:kiss07}
or discrete chemical oscillators~\cite{ref:taylor09,ref:tinsley12,ref:nkomo13}.
On the other hand,
a phase description approach to pattern formation
resulted in the Kuramoto-Sivashinsky equation~\cite{ref:kuramoto84,ref:kuramoto76}
(see also Refs.~\cite{ref:sivashinsky77,ref:kuramoto80,ref:kuramoto84ptp}):
\begin{equation}
  \partial_t \psi(\bd{r}, t)
  = - \nabla^2 \psi - \nabla^4 \psi + (\nabla \psi)^2,
  \label{eq:Kuramoto-Sivashinsky_equation}
\end{equation}
which exhibits spatiotemporal chaos called phase turbulence
\cite{ref:kim01,ref:mikhailov06,ref:mikhailov13}.
In this Letter,
we consider a nonlocal Kuramoto-Sakaguchi model
and derive a Kuramoto-Sivashinsky equation from it.
Namely, we clarify a connection between the above two phase equations.

\paragraph{Derivation.}

We consider a system of nonlocally coupled phase oscillators
described by the following equation~\cite{ref:kuramoto02}:
\begin{equation}
  \partial_t \phi(\bd{r}, t)
  = \omega(\bd{r})
  - \int d\bd{r}'\, G(\bd{r} - \bd{r}')
  \sin\bigl( \phi(\bd{r}, t) - \phi(\bd{r}', t) + \alpha \bigr),
  \label{eq:nonlocal_Kuramoto-Sakaguchi_model}
\end{equation}
where $\phi(\bd{r}, t) \in \mathbb{S}^1$ is the phase at location $\bd{r}$ and time $t$.
The nonlocal coupling function $G(\bd{r})$ is isotropic and normalized as $\int d\bd{r}\, G(\bd{r}) = 1$.
The type of the phase coupling function is assumed to be in-phase coupling, i.e., $|\alpha| < \pi /2$.
We note that this system is heterogeneous owing to the spatially dependent frequency $\omega(\bd{r})$,
which is independently drawn from an identical distribution $g(\omega)$ at each point.
Equation~(\ref{eq:nonlocal_Kuramoto-Sakaguchi_model}) can be called a nonlocal Kuramoto-Sakaguchi model
because of the similarity to Eq.~(\ref{eq:Kuramoto-Sakaguchi_model}).

Introducing a complex order parameter $A(\bd{r}, t)$
with modulus $R(\bd{r}, t)$ and phase $\Theta(\bd{r}, t)$ through
\begin{equation}
  A(\bd{r}, t)
  = R(\bd{r}, t) e^{i \Theta(\bd{r}, t)}
  = \int d\bd{r}'\, G(\bd{r} - \bd{r}') e^{i \phi(\bd{r}', t)},
  \label{eq:order_paramter_phi}
\end{equation}
we rewrite Eq.~(\ref{eq:nonlocal_Kuramoto-Sakaguchi_model}) as
\begin{equation}
  \partial_t \phi(\bd{r}, t)
  = \omega(\bd{r})
  - \frac{1}{2 i} \Bigl( \bar{A}(\bd{r}, t) \, e^{i \phi(\bd{r}, t)} \, e^{i \alpha} - {\rm c.c.} \Bigr),
  \label{eq:nonlocal_Kuramoto-Sakaguchi_model_A}
\end{equation}
where $\bar{A}(\bd{r}, t)$ is the complex conjugate of $A(\bd{r}, t)$.
Applying mean-field theory to Eq.~(\ref{eq:nonlocal_Kuramoto-Sakaguchi_model_A}),
we obtain the following continuity equation:
\begin{align}
  & \frac{\partial}{\partial t} f(\phi, \omega, \bd{r}, t)
  \nonumber \\
  &= - \frac{\partial}{\partial \phi} \Biggl[ \Biggl\{
    \omega - \frac{1}{2 i} \Bigl( \bar{A} \, e^{i \phi} \, e^{i \alpha} - {\rm c.c.} \Bigr)
    \Biggr\} f(\phi, \omega, \bd{r}, t) \Biggr],
  \label{eq:continuity_equation}
\end{align}
where the probability density function $f(\phi, \omega, \bd{r}, t)$
satisfies the following normalization conditions for each location $\bd{r}$ and each time $t$:
\begin{align}
  \int_{-\infty}^{\infty} d\omega \int_0^{2 \pi} d\phi\, f(\phi, \omega, \bd{r}, t)
  &= 1,
  \\
  \int_0^{2 \pi} d\phi \, f(\phi, \omega, \bd{r}, t)
  &= g(\omega),
\end{align}
and the complex order parameter $A(\bd{r}, t)$ is given by
\begin{equation}
  A(\bd{r}, t)
  = \int d\bd{r}'\, G(\bd{r} - \bd{r}') \int_{-\infty}^{\infty} d\omega \int_0^{2 \pi} d\phi\,
  e^{i \phi} f(\phi, \omega, \bd{r}', t).
  \label{eq:order_paramter_f}
\end{equation}
Now, we utilize the Ott-Antonsen ansatz~\cite{ref:ott08,ref:ott09}:
\begin{align}
  & f(\phi, \omega, \bd{r}, t)
  \nonumber \\
  &= \frac{g(\omega)}{2 \pi} \left[ 1 + \sum_{n=1}^{\infty}
    \Biggl\{ \Bigl( a(\omega, \bd{r}, t) \Bigr)^n e^{i n \phi} + {\rm c.c.} \Biggr\} \right].
  \label{eq:Ott-Antonsen_ansatz}
\end{align}
Substituting Eq.~(\ref{eq:Ott-Antonsen_ansatz}) into Eq.~(\ref{eq:continuity_equation}),
we obtain the following equation:
\begin{equation}
  \partial_t a(\omega, \bd{r}, t)
  = - i \omega a - \frac{1}{2} \Bigl( A \, a^2 \, e^{-i \alpha} - \bar{A} \, e^{i \alpha} \Bigr),
  \label{eq:a}
\end{equation}
where the complex order parameter $A(\bd{r}, t)$ is given by
\begin{equation}
  A(\bd{r}, t)
  = \int d\bd{r}'\, G(\bd{r} - \bd{r}') \int_{-\infty}^{\infty} d\omega\,
  g(\omega) \, \bar{a}(\omega, \bd{r}', t).
\end{equation}
In the case of the Lorentzian frequency distribution
\begin{equation}
  g(\omega) = \frac{\gamma}{\pi} \, \frac{1}{(\omega - \omega_0)^2 + \gamma^2},
\end{equation}
the complex order parameter $A(\bd{r}, t)$ is given by
\begin{equation}
  A(\bd{r}, t) = \int d\bd{r}'\, G(\bd{r} - \bd{r}') z(\bd{r}', t),
\end{equation}
where the complex variable $z(\bd{r}, t)$ is defined as
\begin{equation}
  z(\bd{r}, t) = \bar{a}(\omega = \omega_0 - i \gamma, \bd{r}, t).
\end{equation}
From Eq.~(\ref{eq:a}),
we thus obtain the following complex amplitude equation for $z(\bd{r}, t)$ in a closed form:
\begin{equation}
  \partial_t z(\bd{r}, t)
  = (- \gamma + i \omega_0) z - \frac{1}{2} \Bigl( \bar{A} \, z^2 \, e^{i \alpha} - A \, e^{-i \alpha} \Bigr).
  \label{eq:nonlocal_Ott-Antonsen_equation}
\end{equation}
We note that this complex amplitude field is homogeneous.
Equation~(\ref{eq:nonlocal_Ott-Antonsen_equation})
can be called a nonlocal Ott-Antonsen equation,
which was first derived by Laing~\cite{ref:laing09}.
This type of equation has been derived and investigated by several authors
\cite{ref:laing09,ref:laing11,ref:lee11,ref:bordyugov10,ref:wolfrum11,ref:omelchenko13},
but we clarify yet another point mentioned below.

Equation~(\ref{eq:nonlocal_Ott-Antonsen_equation}) can also be written in the following form:
\begin{equation}
  \partial_t z(\bd{r}, t)
  = (\varepsilon + i \Omega_0) z - \beta |z|^2 z
  + \Bigl[ \bar{\beta} \, (A - z) - \beta z^2 (\bar{A} - \bar{z}) \Bigr],
  \label{eq:nonlocal_Ott-Antonsen_equation_coupling_form}
\end{equation}
where the parameters are given by
\begin{equation}
  \varepsilon = \frac{\cos\alpha}{2} - \gamma,
  \qquad
  \Omega_0 = \omega_0 - \frac{\sin\alpha}{2},
  \qquad
  \beta = \frac{1}{2} e^{i\alpha}.
\end{equation}
We note that the first and second terms on the right-hand side
of Eq.~(\ref{eq:nonlocal_Ott-Antonsen_equation_coupling_form})
represent the local dynamics called a Stuart-Landau oscillator~\cite{ref:kuramoto84},
and the remaining term represents the coupling.
From the condition of $\varepsilon > 0$,
collective oscillations exist in the following region:
\begin{equation}
  \cos\alpha > 2 \gamma.
  \label{eq:Hopf}
\end{equation}
Considering the long-wave dynamics of the complex variable $z(\bd{r}, t)$,
we expand the nonlocal coupling term as
\begin{equation}
  A(\bd{r}, t) - z(\bd{r}, t)
  = \sum_{n=1}^\infty G_{2n} \nabla^{2n} z(\bd{r}, t),
  \label{eq:series}
\end{equation}
where $G_{2n}$ is the $2n$-th moment of $G(\bd{r})$.
Substituting Eq.~(\ref{eq:series}) into Eq.~(\ref{eq:nonlocal_Ott-Antonsen_equation_coupling_form}),
we obtain the following equation:
\begin{align}
  \partial_t z(\bd{r}, t)
  = (\varepsilon + i \Omega_0) z - \beta |z|^2 z
  &+ G_2 \Bigl[ \bar{\beta} \, \nabla^2 z - \beta z^2 \nabla^2 \bar{z} \Bigr]
  \nonumber \\
  &+ G_4 \Bigl[ \bar{\beta} \, \nabla^4 z - \beta z^2 \nabla^4 \bar{z} \Bigr]
  \nonumber \\
  &+ O\left( \nabla^6 z \right).
  \label{eq:nonlocal_Ott-Antonsen_equation_diffusion_form}
\end{align}
Owing to the long-wave dynamics of the complex variable $z(\bd{r}, t)$,
the complex order parameter $A(\bd{r}, t)$ is approximated as follows:
\begin{equation}
  A(\bd{r}, t)
  = R(\bd{r}, t) e^{i \Theta(\bd{r}, t)}
  \simeq z(\bd{r}, t).
  \label{eq:A_simeq_z}
\end{equation}
Therefore, the phase of $z(\bd{r}, t)$ can also be considered as the collective phase $\Theta(\bd{r}, t)$.

The uniformly oscillating solution $X_0(\Theta)$ of
Eq.~(\ref{eq:nonlocal_Ott-Antonsen_equation_diffusion_form})
or
Eq.~(\ref{eq:nonlocal_Ott-Antonsen_equation_coupling_form})
is described by
\begin{equation}
  z(\bd{r}, t) = X_0(\Theta) = \sqrt{\frac{\varepsilon}{{\rm Re}\, \beta}} e^{i\Theta},
  \qquad
  \dot{\Theta}(\bd{r}, t) = \Omega,
  \label{eq:uniformly_oscillating_solution}
\end{equation}
where the collective frequency $\Omega$ is obtained as
\begin{equation}
  \Omega
  = \Omega_0 - \varepsilon \frac{{\rm Im}\, \beta}{{\rm Re}\, \beta}
  = \omega_0 - \sin\alpha + \gamma \tan\alpha.
\end{equation}
The left and right Floquet eigenvectors associated with the zero eigenvalue $\Lambda_0 = 0$ are respectively given by
\begin{equation}
  U_0^\ast(\Theta) = i \sqrt{\frac{{\rm Re}\, \beta}{\varepsilon}} \frac{\beta}{{\rm Re}\, \beta} e^{i\Theta},
  \qquad
  U_0(\Theta) = i \sqrt{\frac{\varepsilon}{{\rm Re}\, \beta}} e^{i\Theta},
\end{equation}
where $U_0(\Theta) = dX_0(\Theta) / d\Theta$.
The left and right Floquet eigenvectors associated with another eigenvalue $\Lambda_1 = - 2 \varepsilon$ are respectively given by
\begin{equation}
  U_1^\ast(\Theta) = \sqrt{\frac{{\rm Re}\, \beta}{\varepsilon}} e^{i\Theta},
  \qquad
  U_1(\Theta) = \sqrt{\frac{\varepsilon}{{\rm Re}\, \beta}} \frac{\beta}{{\rm Re}\, \beta} e^{i\Theta}.
\end{equation}
These eigenvectors satisfy the following orthonormalization condition:
\begin{equation}
  {\rm Re}\, \Bigl[ \bar{U}_p^\ast(\Theta) U_q(\Theta) \Bigr] = \delta_{pq},
\end{equation}
where $p, q = 0, 1$.
Although the Floquet eigenvectors and their inner product are expressed by complex numbers
for the sake of convenience in the analytical calculations performed below,
they exactly coincide with the known results for the Stuart-Landau oscillator~\cite{ref:kuramoto84}.

Applying the second-order phase reduction method~\cite{ref:kuramoto84}
to Eq.~(\ref{eq:nonlocal_Ott-Antonsen_equation_diffusion_form}),
we derive the following Kuramoto-Sivashinsky equation for $\Theta(\bd{r}, t)$:
\begin{equation}
  \partial_t \Theta(\bd{r}, t)
  = \Omega
  + \nu \nabla^2 \Theta
  + \mu (\nabla \Theta)^2
  - \lambda \nabla^4 \Theta.
  \label{eq:Kuramoto-Sivashinsky-type_equation}
\end{equation}
Defining the following operator,
\begin{equation}
  \hat{D}_{2n} U(\Theta)
  = G_{2n} \left[ \bar{\beta} \, U(\Theta) - \beta \bigl( X_0(\Theta) \bigr)^2 \bar{U}(\Theta) \right],
\end{equation}
we can write the coefficients as follows.
First, the coefficient $\nu$ is given by
\begin{equation}
  \nu
  = {\rm Re}\, \Bigl[ \bar{U}_0^\ast(\Theta) \hat{D}_2 U_0(\Theta) \Bigr]
  = G_2 \left( \cos\alpha - \frac{\gamma}{\cos^2 \alpha} \right).
  \label{eq:nu}
\end{equation}
Second, the coefficient $\mu$ is given by
\begin{equation}
  \mu
  = {\rm Re}\, \Bigl[ \bar{U}_0^\ast(\Theta) \hat{D}_2 U_0'(\Theta) \Bigr]
  = G_2 \sin\alpha.
  \label{eq:mu}
\end{equation}
Finally, the coefficient $\lambda$ is given by
\begin{align}
  \lambda
  =& \; \Lambda_1^{-1} {\rm Re}\, \Bigl[ \bar{U}_0^\ast(\Theta) \hat{D}_2 U_1(\Theta) \Bigr]
  {\rm Re}\, \Bigl[ \bar{U}_1^\ast(\Theta) \hat{D}_2 U_0(\Theta) \Bigr]
  \nonumber \\
  &- {\rm Re}\, \Bigl[ \bar{U}_0^\ast(\Theta) \hat{D}_4 U_0(\Theta) \Bigr]
  \nonumber \\
  =& \; G_2^2 \left( \frac{1}{\cos\alpha - 2\gamma} \right) \left( \frac{\gamma \tan\alpha}{\cos\alpha} \right)^2
  \nonumber \\
  &- G_4 \left( \cos\alpha - \frac{\gamma}{\cos^2 \alpha} \right).
  \label{eq:lambda}
\end{align}
By introducing the following collective phase gradient, $\bd{V}(\bd{r}, t) = 2 \nabla \Theta(\bd{r}, t)$,
Eq.~(\ref{eq:Kuramoto-Sivashinsky-type_equation}) is also written as
\begin{equation}
  \partial_t \bd{V}(\bd{r}, t)
  = \nu \nabla^2 \bd{V}
  + \mu \bd{V} \cdot \nabla \bd{V}
  - \lambda \nabla^4 \bd{V},
  \label{eq:Kuramoto-Sivashinsky-type_equation_V}
\end{equation}
which is also called the Kuramoto-Sivashinsky equation.

Here, we note the derivation of the coefficients in another manner~\cite{ref:kuramoto84ptp}.
The linear stability analysis of Eq.~(\ref{eq:nonlocal_Ott-Antonsen_equation_coupling_form})
around the uniformly oscillating solution $X_0(\Theta)$ given in Eq.~(\ref{eq:uniformly_oscillating_solution})
provides two linear dispersion curves $\Lambda_{\pm}(q)$:
one is the phase branch, which satisfies $\Lambda_{+}(0) = \Lambda_0 = 0$;
the other is the amplitude branch, which satisfies $\Lambda_{-}(0) = \Lambda_1 = -2\varepsilon$.
The phase branch is expanded with respect to the wave number $q$ as follows:
$\Lambda_{+}(q) = - \nu q^2 - \lambda q^4 + O(q^6)$.
The linear coefficients, $\nu$ and $\lambda$, are also obtained in this way.
In addition, the coefficient $\mu$ is found from the dependence of the frequency $\Omega_k$ on the wave number $k$
for plane wave solutions to Eq.~(\ref{eq:nonlocal_Ott-Antonsen_equation_coupling_form}) as follows:
$\Omega_k = \Omega + \mu k^2 + O(k^4)$.

\begin{figure*}
  \begin{minipage}{0.45\hsize}
    \begin{center}
      \includegraphics[width=\hsize,clip]{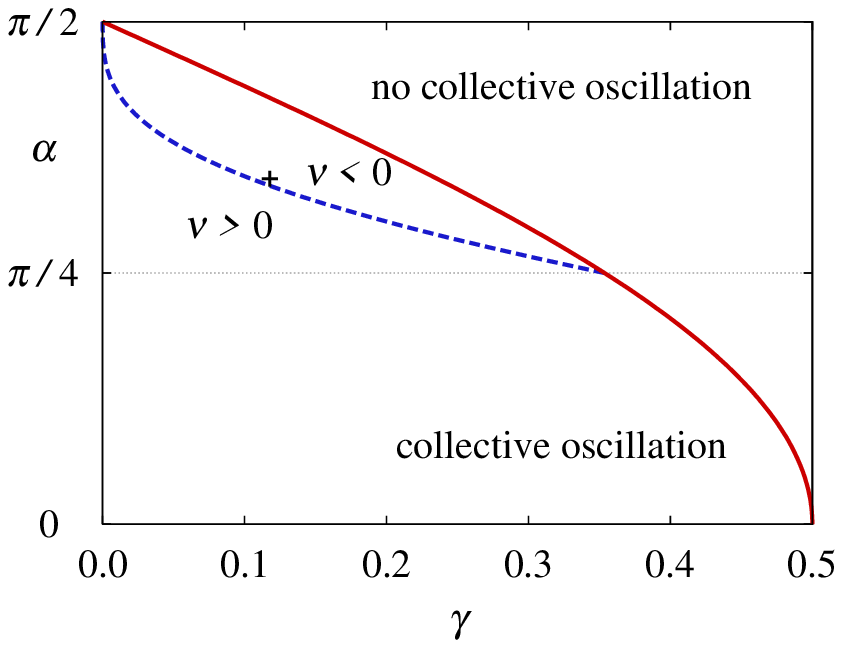}
      \caption{(color online).
        Phase diagram in the parameter plane, $\alpha$ and $\gamma$.
        The red solid curve indicates the Hopf bifurcation line, $\gamma = \cos(\alpha)/2$.
        The blue broken curve indicates the transition line to turbulence, $\gamma = \cos^3(\alpha)$.
        The dotted line ($\alpha = \pi/4$) is a guide for the eye.
        The plus sign ($+$) indicates $\alpha = 1.08$ and $\gamma = \cos(\alpha)/4$,
        which were used in all the numerical simulations.
      }
      \label{fig:1}
    \end{center}
  \end{minipage}
  \begin{minipage}{0.05\hsize} \quad \end{minipage}
  \begin{minipage}{0.45\hsize}
    \begin{center}
      \includegraphics[width=\hsize,clip]{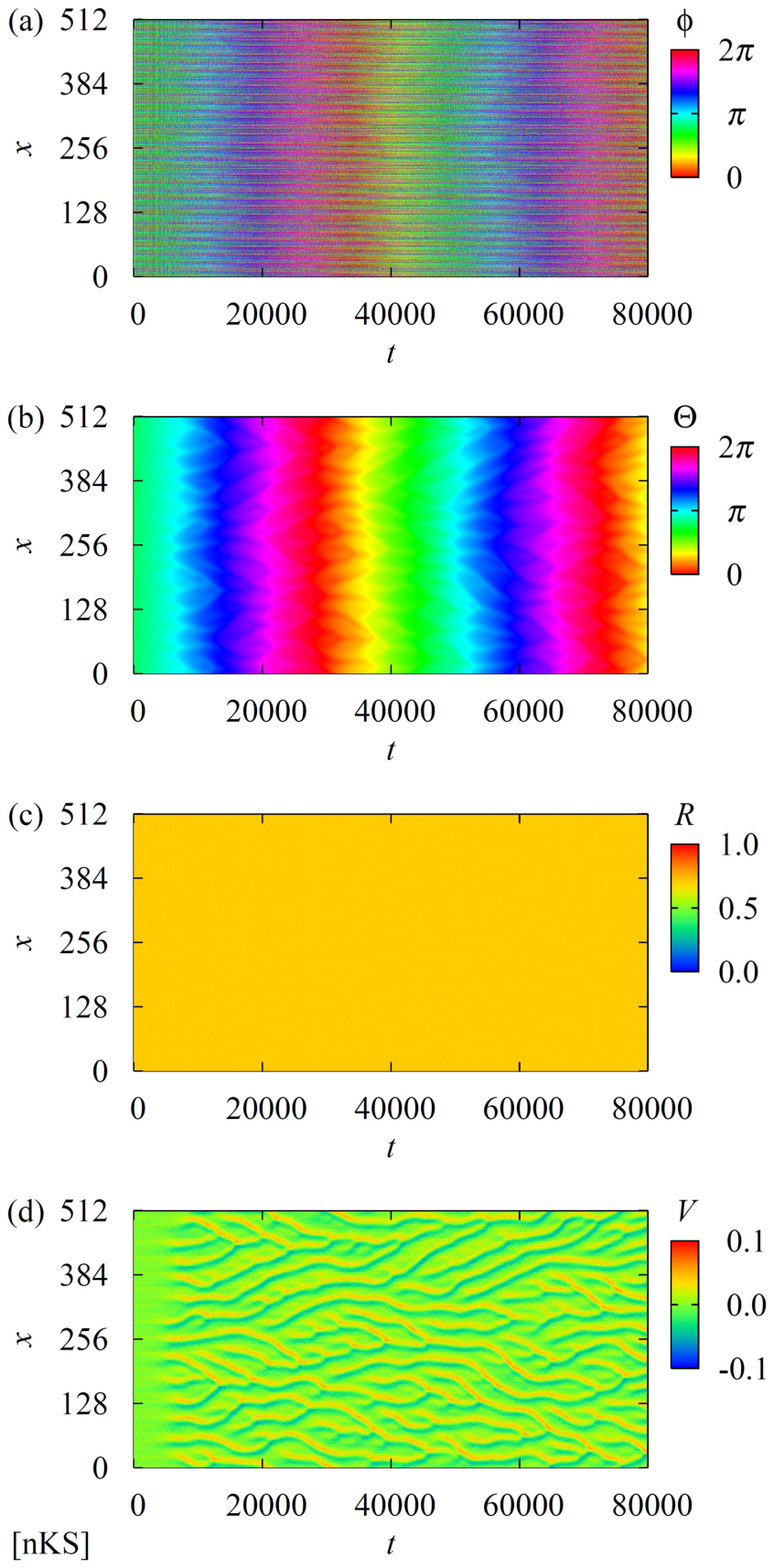}
      \caption{(color online).
        Numerical simulation of the nonlocal Kuramoto-Sakaguchi (nKS) model.
        The number of grid points is $2^{19}$.
        (a) Local phase $\phi(x,t)$.
        (b) Order parameter phase $\Theta(x,t)$.
        (c) Order parameter modulus $R(x,t)$.
        (d) Order parameter phase gradient $V(x,t)$.
      }
      \label{fig:2}
    \end{center}
  \end{minipage}
\end{figure*}

We also note that the collective phase diffusion coefficient $\nu$ can be negative 
despite the in-phase coupling, $|\alpha| < \pi/2$,
and that negative diffusion results in spatiotemporal chaos.
In fact, from the condition of $\nu < 0$,
spatiotemporal chaos can occur in the following region:
\begin{equation}
  \cos^3\alpha < \gamma.
  \label{eq:transition}
\end{equation}
At the onset of collective oscillation, i.e., $\cos\alpha = 2\gamma$,
spatiotemporal chaos can occur in the following region:
\begin{equation}
  \cos^2\alpha < \frac{1}{2},
  \label{eq:transition_onset}
\end{equation}
which gives $|\alpha| > \pi/4$.
Figure~\ref{fig:1} shows the phase diagram,
which is composed of Eqs.~(\ref{eq:Hopf}), (\ref{eq:transition}), and (\ref{eq:transition_onset})
in the parameter plane, $\alpha$ and $\gamma$.
When $\pi/4 < |\alpha| < \pi/2$,
the sign of the coefficient $\nu$ changes from positive to negative
as the dispersion parameter $\gamma$ increases;
this transition phenomenon can be called heterogeneity-induced turbulence.

\paragraph{Simulation.}

For the sake of simplicity,
we carried out numerical simulations in one spatial dimension.
As the nonlocal coupling function $G(x)$,
we used the Helmholtz-type Green's function:
\begin{equation}
  G(x)
  = \frac{1}{2\pi} \int_{-\infty}^{\infty} dq \, \frac{e^{i q x}}{1 + q^2}
  = \frac{1}{2} e^{-|x|},
\end{equation}
which gives $G_{2n} = 1$.
The truncation of the Kuramoto-Sivashinsky equation~(\ref{eq:Kuramoto-Sivashinsky-type_equation}) holds
if and only if the collective phase diffusion coefficient $\nu$ is small and negative.
The parameter values are thus chosen to be
$\alpha = 1.08$ and
$\gamma = \cos(\alpha) /4$,
which give
$\nu     \simeq -0.06$,
$\mu     \simeq  0.88$, and
$\lambda \simeq  0.99$.
The central value of the frequency distribution is fixed at
$\omega_0 = \sin(\alpha) - \gamma \tan(\alpha)$,
which gives
$\Omega = 0$.
The system size is $512$, and a periodic boundary condition is imposed.

\begin{figure*}
  \begin{minipage}{0.45\hsize}
    \begin{center}
      \includegraphics[width=\hsize,clip]{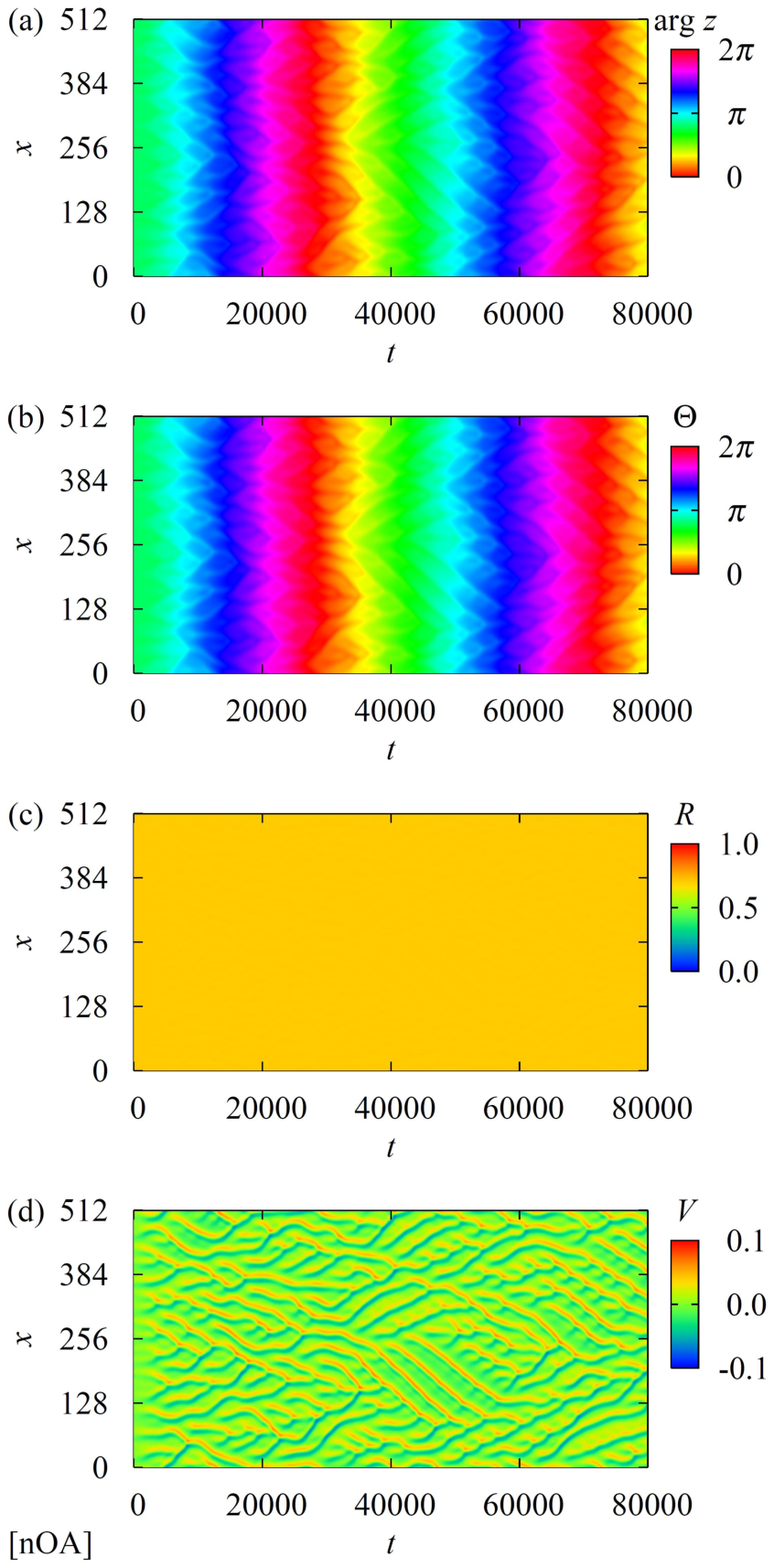}
      \caption{(color online).
        Numerical simulation of the nonlocal Ott-Antonsen (nOA) equation.
        The number of grid points is $2^{11}$.
        (a) Local variable argument $\arg z(x, t)$.
        (b) Order parameter phase $\Theta(x,t) = \arg A(x, t)$.
        (c) Order parameter modulus $R(x,t) = | A(x, t) |$.
        (d) Order parameter phase gradient $V(x,t) = 2 \partial_x \Theta(x,t)$.
      }
      \label{fig:3}
    \end{center}
  \end{minipage}
  \begin{minipage}{0.05\hsize} \quad \end{minipage}
  \begin{minipage}{0.45\hsize}
    \begin{center}
      \includegraphics[width=\hsize,clip]{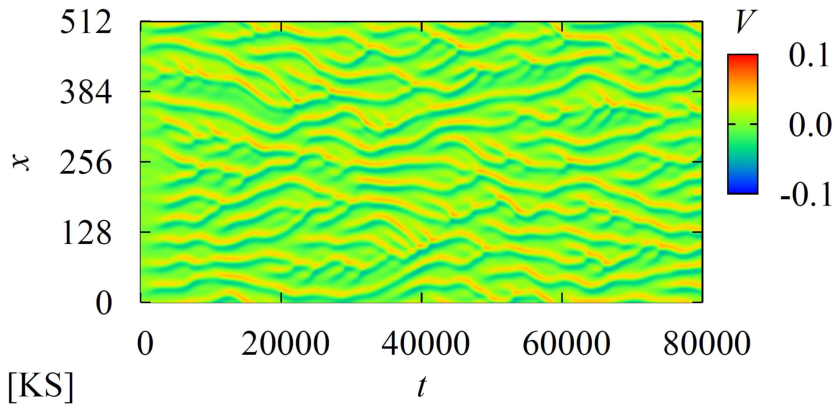}
      \caption{(color online).
        Numerical simulation of the Kuramoto-Sivashinsky (KS) equation.
        The number of grid points is $2^{11}$.
      }
      \label{fig:4}
    \end{center}
  \end{minipage}
\end{figure*}

Numerical simulations of
the nonlocal Kuramoto-Sakaguchi model~(\ref{eq:nonlocal_Kuramoto-Sakaguchi_model})
\footnote{
  This simulation requires a careful setup to suppress finite-size effects.
  The frequency distribution within the characteristic coupling width at each point
  should be Lorentzian as well as possible,
  and each distribution at each point should be the same as each other.
  In our numerical simulation,
  the system size is $2^9$,
  the characteristic coupling width is $2^1$,
  and
  the number of grid points is $2^{19}$.
  Therefore,
  the number of grid points within the characteristic coupling width is $2^{11}$.
  In addition,
  the central value of the frequency distribution
  is compensated with some value of order $10^{-4}$
  to accurately give $\Omega = 0$;
  however, this point is not essential for phase turbulence itself.
},
the nonlocal Ott-Antonsen equation~(\ref{eq:nonlocal_Ott-Antonsen_equation}), and
the Kuramoto-Sivashinsky equation~(\ref{eq:Kuramoto-Sivashinsky-type_equation})
are shown in Figs.~\ref{fig:2}, \ref{fig:3}, and \ref{fig:4}, respectively.
The spatiotemporal evolutions of the collective phase gradient $V(x, t)$
shown in Figs.~\ref{fig:2}(d), \ref{fig:3}(d), and \ref{fig:4}
are remarkably similar to each other.
Equivalently, the spatiotemporal evolutions of the collective phase $\Theta(x, t)$
shown in Figs.~\ref{fig:2}(b) and \ref{fig:3}(b)
are also similar to each other.
As seen in Figs.~\ref{fig:2}(a) and \ref{fig:2}(b),
the spatial pattern of the individual phase $\phi(x, t)$ is non-smooth,
but that of the collective phase $\Theta(x, t)$ is smooth.
As seen in Figs.~\ref{fig:3}(a) and \ref{fig:3}(b),
the phase of $z(x, t)$ can be considered as the collective phase $\Theta(x, t)$,
namely, Eq.~(\ref{eq:A_simeq_z}) is actually valid.
As seen in Figs.~\ref{fig:2}(c) and \ref{fig:3}(c),
the order parameter modulus $R(x, t)$ is almost constant,
so that the phase reduction approximation is also valid.
We thus conclude that this spatiotemporal chaos is
the so-called phase turbulence in the complex order parameter field.
Here, we note that amplitude turbulence also occurs in the large negative $\nu$ region.

\paragraph{Discussion.}

In this Letter,
we studied heterogeneity-induced turbulence in nonlocally coupled oscillators
(i.e., the nonlocal Kuramoto-Sakaguchi model),
where the Kuramoto-Sivashinsky equation was derived
via the nonlocal Ott-Antonsen equation.
In Ref.~\cite{ref:kawamura07},
we have studied noise-induced turbulence in nonlocally coupled oscillators
(i.e., a nonlocal noisy Kuramoto-Sakaguchi model),
where the Kuramoto-Sivashinsky equation has been derived
via the complex Ginzburg-Landau equation~\cite{ref:kuramoto84,ref:kuramoto74,ref:aranson02}
or the nonlinear Fokker-Planck equation~\cite{ref:kuramoto84}.
There exists a remarkable connection
between the nonlocal Kuramoto-Sakaguchi model and the Kuramoto-Sivashinsky equation.

As mentioned in Ref.~\cite{ref:kawamura10a},
noise-induced turbulence in a system of nonlocally coupled oscillators~\cite{ref:kawamura07}
is closely related to
noise-induced anti-phase synchronization
between two interacting groups of globally coupled oscillators~\cite{ref:kawamura10a}.
Similarly,
heterogeneity-induced turbulence in a system of nonlocally coupled oscillators
is also closely related to
heterogeneity-induced anti-phase synchronization
between two interacting groups of globally coupled oscillators~\cite{ref:kawamura10b};
namely, Eqs.~(\ref{eq:nu}) and (\ref{eq:mu}) in this Letter
correspond to Eq.~(31) in Ref.~\cite{ref:kawamura10b}.

In summary,
we derived the Kuramoto-Sivashinsky equation
from the nonlocal Kuramoto-Sakaguchi model
and demonstrated heterogeneity-induced phase turbulence.
We hope that the connection between these two landmark phase equations
will facilitate the theoretical analysis of coupled oscillators
and that heterogeneity-induced turbulence
will be experimentally confirmed in the near future.


\begin{acknowledgments}
  The author is grateful to Yoshiki Kuramoto and Hiroya Nakao for valuable discussions.
  This work was supported by JSPS KAKENHI Grant Number 25800222.
\end{acknowledgments}


\clearpage

\end{document}